# Tunable Terahertz Detection and Generation using FETs operating in the saturation regime.

Tamer Elkhatib,[1,2, a]

[1]*Rensselaer polytechnic Institute, 110, 8th street, Troy, NY, 12180, USA.*
[2]*Dep. of Eng. Mathematics & Physics, Faculty of Engineering, Cairo University, Giza, 12613, EGYPT.*

I report on the experimental observation of DC instability and self-amplification through stimulated emission of 0.2 and 1.63 THz radiation using InGaAs/GaAs HEMT operating in the deep saturation regime at room temperature. I demonstrate both theoretically and experimentally, that the Sub-THz and THz response of FETs are attributable to the rectification of the nonlinear dependence of the device's current-voltage characteristics. FETs function as nonlinear THz mixers and rectifiers, with their open-drain responsivity described by an expression analogous to that of a zero-bias Schottky diode detector. However, operating FETs in the deep saturation regime permits precise tuning of the device to the quantum localized resonance condition and the negative resistance mode at room temperature. Consequently, FETs can be adjusted in the deep saturation regime to facilitate tunable sub-THz and THz detection and generation as well as tunable sub-THz and THz lasing effect. These results are anticipated to significantly impact technological advancements across various fields in the near future.

Field-effect transistors (FETs) and high-electron-mobility transistors (HEMTs) across various technologies and material systems have been extensively investigated for terahertz detection over the last few decades. The study of plasma oscillation phenomena within field-effect transistor channels commenced with the theoretical prediction[1] that these waves may exhibit one of many different dispersion relations. Subsequent research observed infrared absorption[2] and weak infrared emission[3] associated with the excitation of plasma oscillations in silicon inversion layers of metal-oxide-semiconductor FETs (MOSFETs).

In recent decades, interest in plasma phenomena within FET channels has intensified due to the possibility of utilizing these plasma oscillations in the tunable emission and detection of THz radiation [4,5]. The widely spread Dyakonov-Shur hydrodynamic model[5] highlighted that an instability of direct current flows through the device channel can occur when AC resonance oscillations takes place across the entire transistor's channel, resulting in the emission of terahertz radiation, while the device's nonlinearity facilitates detection and mixing in the terahertz range beyond the device's cutoff frequency. Previously, this quantum wave DC instability has only been observed at very low temperatures, and most research studies have been confined to open-drain or unbiased transistors due to the limitations of the Dyakonov-Shur physical model to this zero-drain biased operational mode.

The academic research in this field of THz electronics[6-26] has led to resonant and broadband (non-resonant) detection or emission of sub-THz and THz radiation by nanoscale FETs. Though the THz emitters are still rather weak, the THz detectors already show good performance: they are tunable[6-26] by varying the gate voltage and the drain current, with fast response time[21] and demonstrate relatively low noise equivalent power (NEP) up to room temperature[18]. Hence, they are promising candidates for terahertz systems applications.

As demonstrated in Refs. [6] and [17], the efficiency of broadband detection can be significantly enhanced by applying the drain current $j$ to drive the FETs into saturation regime. The physical model[17] indicates that the response significantly increases as $j$ approaches the saturation current $j_{sat}$; (knee current on the I-V curve). However, the terahertz (THz) response of FETs operating in the deep saturation regime $j>j_{sat}$ has not been explored either experimentally or theoretically until the work published in Ref. [27].

I start with a theoretical discussion. At very low drain biasing voltages, the I-V characteristics of the transistor can be expressed as $j_d = G(U_{gs}, U_{gd}) U_{ds}$, where $G = (R_{ch})^{-1}$, is the channel conductivity dependent on both $U_{gs}$ the gate-to-source and $U_{gd}$ gate-to-drain voltages. The incident terahertz radiation couples to the transistor channel from both the source and drain sides. Consequently, an induced direct current will flow in the device such that: $\delta J_d = (G + \delta G)\delta U_{ds}$. If an external load resistor $R_L$ is connected to the transistor, the resulting DC change is given by: $\delta j_d = -\delta U_{ds}/R_L = \delta U_{ds}/R_{ch} + (\delta G)\delta U_{ds}$, where the change in conductivity can be expressed as a function of effective variations in the applied $U_{gs}$ and $U_{gd}$ as: $\delta G = 0.5 \times \delta U_{gs} \times \partial G/\partial U_{gs} + 0.5 \times \delta U_{gd} \times \partial G/\partial U_{gd}$.

The radiation-induced drain-to-source voltage can also be expressed as: $\delta U_{ds} = \delta U_{gs} - \delta U_{gd}$. Given that the device operates in the open-drain condition with very small signal variations at both the source and drain terminals, we can assume that $\partial G/\partial U_{gs} = \partial G/\partial U_{gd}$, and hence the induced DC terahertz response in the device can be expressed as:

[a] Author to whom correspondence should be addressed: Electronic mail: t.a.elkhatib@cu.edu.eg

$$\delta U_{ds} = -\frac{1}{2}\left(\delta U_{gs}^2 - \delta U_{gd}^2\right) \times \frac{\partial G}{\partial U_{gs}} \times \frac{R_L \times R_{ch}}{R_L + R_{ch}} \quad (1)$$

Two radiation-induced alternating current (AC) voltage sources: $U_a \cos\omega t$ and $U_b \cos\omega t$ can be considered at the transistor's source and drain terminals, respectively. Consequently, the effective direct current (DC) variations in the gate-to-source and gate-to-drain voltages are represented by: $\delta U_{gs}^2 = U_a^2/2$ and $\delta U_{gd}^2 = U_b^2/2$ and Equation (1) elucidates why the response can be either positive or negative, contingent upon the ratio of $U_a^2/U_b^2$, which is primarily influenced by the metallization layout of the transistor terminals. In the context of open-drain detection, the coupling of terahertz radiation is compromised unless special attention is devoted to the design of the transistor layout.

By disregarding the external load effect in Equation (1), the intrinsic terahertz (THz) responsivity of the open-drain transistor can be expressed as a function of the device's current-voltage (I-V) characteristics, analogous to that of a zero-bias Schottky diode detector. For a transistor, the THz responsivity is given by:

$$R \propto \pm \frac{\partial^2 I}{\partial U_{gs} \partial U_{ds}} \frac{\partial U_{ds}}{\partial I}\bigg|_{I=0} \quad (2)$$

For a Schottky diode detector, it is expressed as:

$$R \propto -\frac{\partial^2 I}{\partial U^2} \frac{\partial U}{\partial I}\bigg|_{I=0} \quad (3)$$

This confirms that the mechanisms of terahertz detection in both devices are identical due to the rectification of the nonlinear behavior of their current-voltage characteristics. However, transistors possess the advantage of having their detectivity adjustable through the application of gate voltage. It is important to note that the THz response of FETs, as described by Eq. (1), is applicable for gate-bias voltages both above and below the transistor threshold voltage $U_{th}$. In the regime above the threshold, it can be derived from the drain current equation that: $R_{ch} \times \partial G/\partial U_{gs} = 1/(U_{gs} - U_{th})$. In such a scenario, an approximate behavioral expression for the response can be formulated as:

$$\Delta U_{ds} = -\frac{1}{2}\left(\frac{U_a^2}{2} - \frac{U_b^2}{2}\right)\frac{1}{(U_{gs}-U_{th})}\frac{R_L}{R_L+R_{ch}} \quad (4)$$

In the subthreshold regime, the transistor drain current is proportional to $\exp(U_{gs}/\eta U_t)$ and $\exp(U_{ds}/\eta U_t)$, where $\eta$ represents the subthreshold ideality factor and $U_t$ denotes the thermal voltage. It can be easily found out the following expression: $R_{ch} \times \partial G/\partial U_{gs} = 1/\eta U_t$, and an additional simplified behavioral equation for the terahertz response at gate biasing below the threshold voltage can be derived as:

$$\Delta U_{ds} = -\frac{1}{2}\left(\frac{U_a^2}{2} - \frac{U_b^2}{2}\right)\frac{1}{\eta U_t}\frac{R_L}{R_L+R_{ch}} \quad (5)$$

In the context of drain-biased transistors, the voltage of the induced terahertz (THz) radiation across the drain and source terminals can be approximately formulated as follows: $\delta U_{ds/biased} = \delta U_{gs/biased} \times \frac{\partial U_{ds}}{\partial U_{gs}} \times \frac{R_L}{R_L+R_{ch}}$. Consequently, the simple approximate behavioral expression for the induced direct current (DC) terahertz response, utilizing the transistor's transconductance $G_m$, can be formulated as:

$$\delta U_{ds/biased} = -\delta U_{gs/b} \times G_m \times \frac{R_{ch} \times R_L}{R_{ch}+R_L} \quad (6)$$

$$\Delta U_{ds/biased} = -\frac{U_a^2}{2} \times \frac{1}{2}\frac{\partial G_m}{\partial U_{gs}} \times \frac{R_{ch} \times R_L}{R_{ch}+R_L} \quad (7)$$

In this study, I demonstrated that for $j>j_{sat}$, a significant enhancement in the broadband terahertz response can be achieved, the response continues to increase linearly with the dc drain voltage (or dc drain current), potentially reaching substantial values. The phenomenon of THz rectification in the deep saturation operation regime holds considerable significance for THz electronics and serves as a pivotal solution for tuning the device into the negative resistance resonance mode or, in other terms, achieving the sub-THz and THz lasing effect at room temperature. I assert that the findings of this research have the potential to be transformative.

To elucidate how self-amplification by stimulated emission at sub-THz and THz frequencies can be observed in FET channels, consider the following analogy: the standing wave resonance of a sound wave in a tube is determined by three primary parameters: the frequency of the wave, the length of the tube, and the density of particles within the tube. Regrettably, researchers in the field of THz electronics have historically concentrated their efforts predominantly on the open-drain FETs operation regime. In the zero-drain-biased mode, the conductivity of the transistor channel is nearly uniform, allowing the hydrodynamic equations from device physics (Euler and Continuity) to be solved, yielding a simplified closed-solution formula.

Conversely, in the drain-biased FETs, channel conductivity is non-uniform with part that is fully depleted from charge carriers, and the hydrodynamic equations cannot be solved analytically; only a numerical solver can provide a solution without a closed-form formula. The primary challenge in achieving THz plasma resonance in the open-drain operation mode is that the channel length is invariably fixed and equivalent to the physical distance of the transistor gate[4,5], this requires to lock the input THz Ac signal at the exact value to match the localized resonance oscillations in the transistor channel, which is experimentally non-practical. Currently, fine tuning the frequency of commercially available THz sources remains challenging; In addition, the transistor channel demonstrates strong resistive part within the context of high frequency Drude model, which only allows damped THz Ac oscillations within the transistor channel (exponentially decaying evanescent oscillations).

Some researchers have concentrated on the very low operating temperatures to nearly freeze the transistor channel, thereby reducing and tuning its channel resistance to lower values below the critical temperature of superconductivity, which can facilitate localized resonance oscillations with relatively longer channel devices. They employed a cascaded design to mitigate the weak resonance mode by amplifying the output THz radiation to the greatest extent possible. However, the extremely low temperatures required for quantum cascade THz lasers[28] restrict their applicability in mass volume technology market and only a few limited terahertz applications are possible. The primary solution proposed in this research, which enables room temperature tunable detection and generation, as well as lasing effect across sub-THz and THz frequencies, involves operating the FETs and HEMTs in the deep saturation regime. This saturation operation regime is fully independent of any transistor material or bandgap limits.

As illustrated in Fig. 1, the effective resistive channel length is adjusted by varying the applied drain-to-source voltage or drain current, allowing localized resonance oscillations mode or the negative resistance mode to occur independently of any transistor gate length $L$.

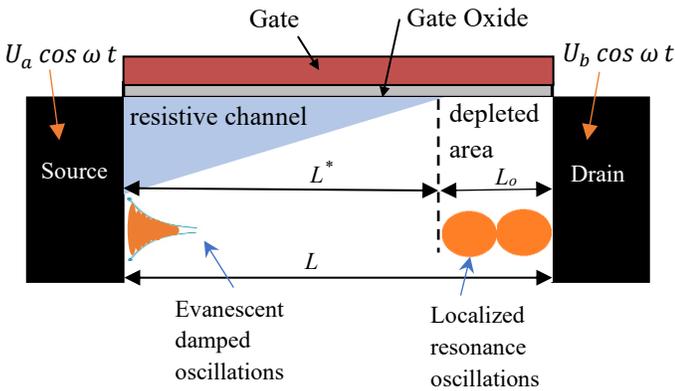

FIG. 1 (color online) Schematic of the channel of field effect transistor operating in the deep saturation regime. The localized resonance oscillations at the drain side happen when the effective resistive channel length $L^*$ (depleted channel length $L_o$) is tuned to match the incident THz radiation frequency, while the Evanescent damped oscillations happen simultaneously at the source side.

Next, I present an analysis of my experimental data. I employed 0.5$\mu m$ enhancement mode InGaAs/GaAs pseudomorphic HEMTs fabricated by TriQuint Semiconductor as terahertz detectors, all its important device parameters are extracted in Ref. 26. The optically pumped far-infrared gas laser (Coherent SIFIR-50) was utilized as a terahertz source operating at 1.63THz. The laser beam was focused on the transistor using polyethylene lens with focal length of 16mm and modulated by an optical chopper with a fixed frequency of 50 Hz. The induced dc drain voltage $\Delta U_{ds}$ superimposed on the drain-to-source voltage $U_{ds}$ was measured using the lock-in amplifier SR-830.

Fig.2 presents the responsivity curves for both the open-drain and drain-biased configurations Initially, the gate-to-source bias was adjusted for the open-drain scenario across various load resistances, and the results were compared with the simplified physical model described in Eq. (1). Subsequently, the drain-to-source current was tuned while maintaining a fixed gate bias and a constant load resistance of $R_L=10K\ Ohm$, and these findings were compared with Eq. (7).

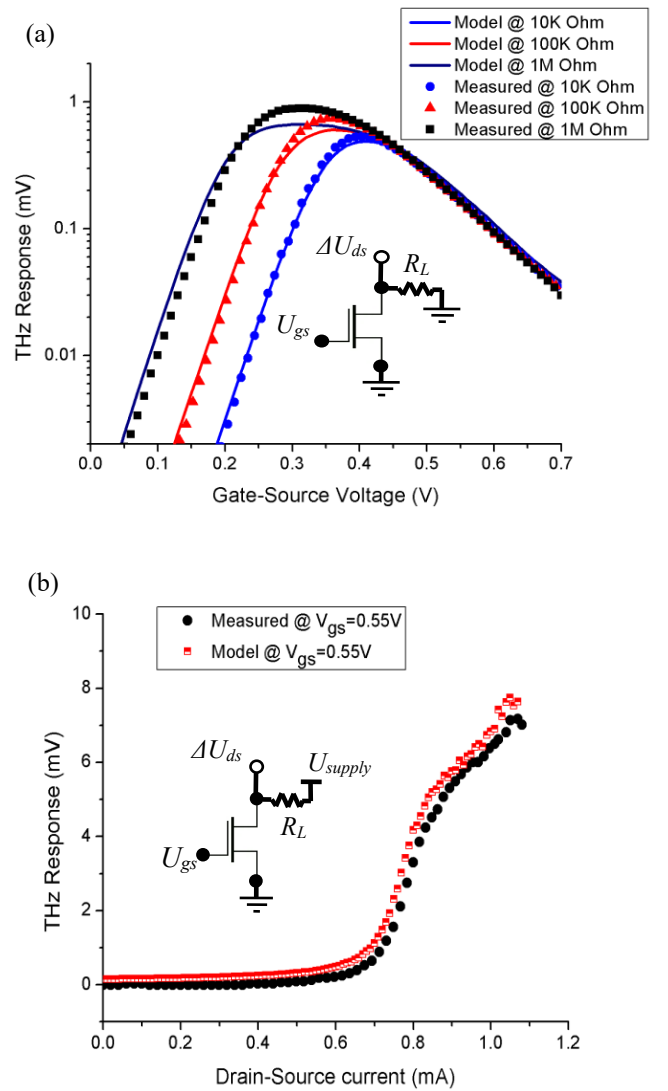

FIG. 2 (color online) Measured and calculated 1.63THz response as a function of: (a) gate-biases for the open-drain case in a semi-log scale, (b) drain-currents for the drain-biased case with a load resistance of *10K Ohm*. The circuit diagrams for each case are shown within the responsivity curves.

It is noteworthy that the aforementioned results exclusively demonstrate the nonlinear rectification of THz radiation at the source and drain terminals of transistors at frequencies significantly exceeding the cutoff frequency. The absence of resonance conditions can be attributed to the mismatch between the device's length and conductivity and the higher frequency of 1.63 THz. I propose an alternative approach to operating the device at extremely low temperatures. By operating the transistor in the subthreshold region, it is feasible to reduce the channel conductivity to lower values, thereby enabling the FET under test to be driven into both the sub-threshold and deep saturation regimes simultaneously.

Consequently, it is possible to adjust the resistive channel length and conductivity to achieve resonance conditions at very high-frequency radiation at room temperature. Figure 3 presents the first experimental observation of DC instability in transistor channel, negative resistance mode, or THz localized resonance oscillations condition in a single long FET (0.5μm technology) device above one THz at room temperature

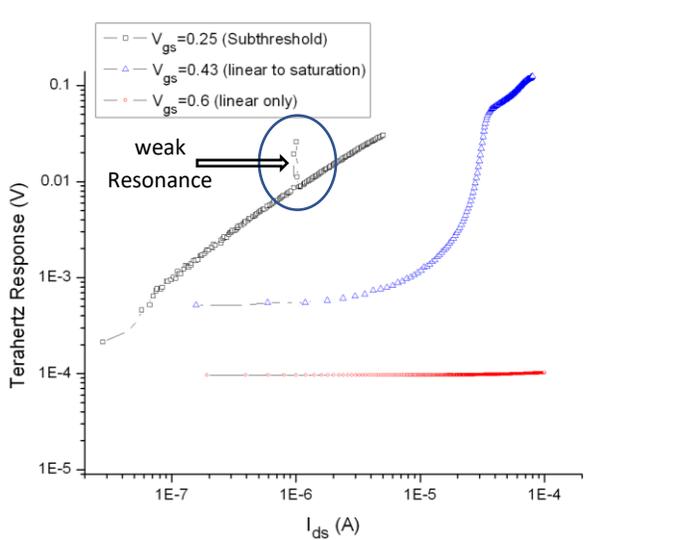

**FIG. 3** (color online) Measured 1.63THz response as a function of drain-currents for the drain-biased case with a load resistance of 100K Ohm in a log-log plot. The first ever observed room temperature weak DC instability or negative resistance operation mode appears using InGaAs/GaAs 0.5um HEMTs at a sub-threshold gate-to-source bias of 0.25V and drain current of about 1μA deep in the saturation regime

To substantiate the pronounced DC instability at room temperature in a long channel transistor device, I conducted experiments using the same fabricated HEMTs under the exposure of a focused 200 GHz radiation beam, generated by a commercial Gunn diode oscillator (refer to Fig. 4). The optical chopper was configured to a modulation frequency of 50 Hz. A computerized 3D NanoMax stage was employed to position the FET device optimally within the focused area of the radiation beam. The THz responsivity at 200 GHz was observed to be at least an order of magnitude greater than that at 1.63 THz, reaching up to 170 V/W.

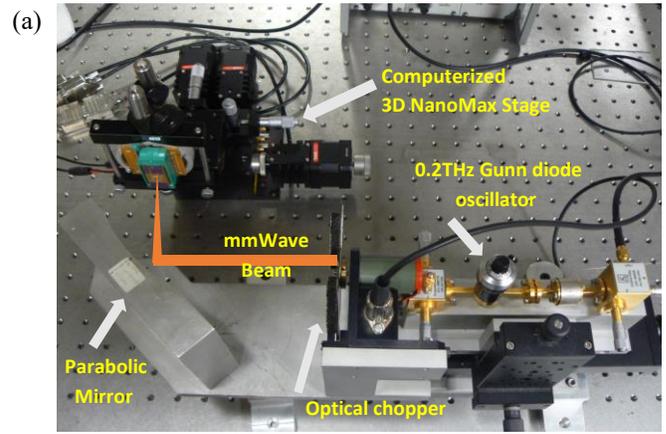

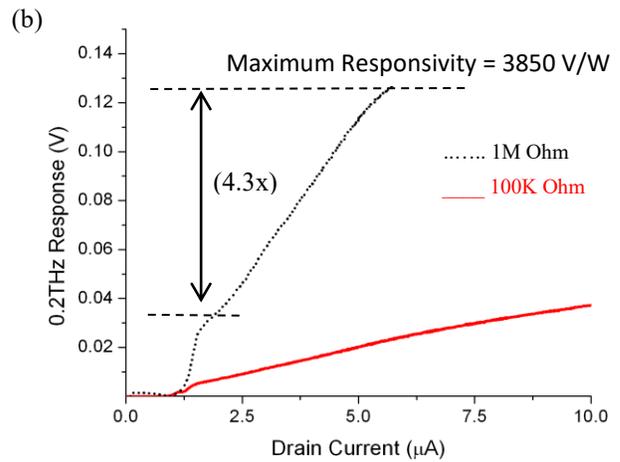

**FIG. 4** (color online) (a) Experimental setup for studying the 200GHz millimeter wave direct coupling with the designed transistors test structure. (b) Measured 0.2THz response as a function of drain-currents for different load resistances. The HEMT transistor is biased in the sub-threshold regime with gate-to-source voltage of 0.25V.

Notably, I report a responsivity as high as 3850 V/W at 200 GHz without the necessity of an integrated coupling antenna, whereas the current waveguide-based Schottky diode detector from Virginia Diodes, Inc[29] exhibits a THz responsivity of 4000 V/W.

The nonlinearity in the 0.2 THz rectification response, with an emphasis on the linear increase in the transistor's response, is clearly depicted in Fig. 4(b) within the deep saturation regime. It is evident that THz responsivity can be enhanced by a factor of 2-4 when transistors operate further into the saturation regime, and additional responsivity gains may be achieved when multiple transistors are connected in series[26]. Furthermore, it is noteworthy that the deep saturation regime can be utilized for correlated imaging of incident THz radiation at sub-wavelength resolution (up to 1/10 of the radiation wavelength), as demonstrated in Refs. [25].

I acknowledge that the noise equivalent power (NEP) in the deep saturation operation regime may be significantly higher than that of the Schottky diode detector due to the elevated drain current in the device, in addition to active rectifications with high gain. Nonetheless, several strategies exist to mitigate the NEP of FETs power THz detectors[26] operating in the deep saturation regime. One of these strategies is the electronic modulation of incident THz radiation above the 1/f corner frequency which can significantly improve the SNR.

In conjunction with the high 0.2 THz responsivity previously mentioned, I successfully adjusted both the gate-to-source and drain-to-source biasing conditions to observe pronounced DC instability at 200 GHz at room temperature. Figure 5 presents, to the best of my knowledge, the first-ever observation of multiple DC instabilities and self-amplification by stimulated emission of 200 GHz radiation. Although no specific analytical solution exists to describe the DC instability under this non-uniform distributed channel resistivity, numerical simulations can address and justify these experimental physics observations. These experimental results were corroborated by numerous samples, consistently yielding the same outcomes.

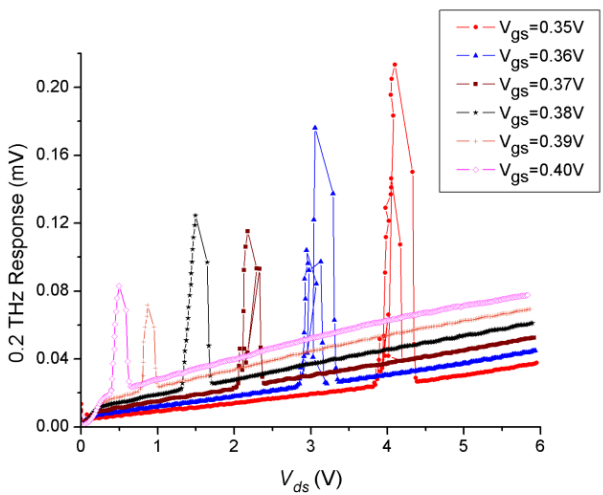

**FIG. 5** (color online) Measured 0.2 THz response as a function of drain-to-source voltages for a load resistance of 100K Ohm. The instability of carrier oscillations and self-amplifications by stimulated emission appears clearly as unexpected peaks at different gate-to-source voltages and drain-to-source biasing.

One can easily analyze the THz response in Fig. 5 and come up with the following indications: For the highest gate-to-source values, the dc instability reaches smaller peaks at minimum drain to source voltage, where the depletion region length $L_o$ is the minimum and the current density in the depleted region is the highest value. On the other hand, the non-linearity rectification response is the best due to the strong amplification factor at higher $U_{gs}$ values. Oppositely, for the smallest gate to source values even below the threshold voltages, the current density in the depletion region is the minimum and the dc instability occurs with highest peaks and with clear hysteresis response at the longest depletion region length $L_o$ with the highest drain to gate voltages.

One can conclude also that this negative resistance instability can be tuned for different Sub-THz frequencies, like 208GHz and 196GHz, very easily because the effective channel length $L^*$ can be simply tuned by $U_{ds}$ and the load resistor. The operation of the FETs and HEMTs as tunable THz detector and emitters in the deep saturation regime opened the opportunity for the world to the room temperature fine tuning of transistors at any frequency band independent from transistor material or bandgap. In another words, transistors can be tuned for frequencies in ultrasound, microwave, millimeter wave, terahertz, far-infrared, visible light, etc. Finally, I should mention that using the nonlinearity of FETs and HEMTs is not limited to a single transistor device. Many integrated circuit solutions such as a one-dimensional array, as a two-dimensional array, or as series connected devices for rectification, mixing, power detection, or high frequency radiation imaging or sensing are also possible.

However, tuning the operation of a single FET in the negative resistance, quantum wave, mode will always demonstrate instability when it is fine-tuned by both gate-to-source and drain current and external load resistor value. Therefore, I have tested the differential mode FETs structure as the one I have designed and tested[25] to stabilize the terahertz AC induced DC instability and obtain stable Sub-THz and THz sources that can operate very well with high power and high frequencies at room temperatures and at both higher and lower temperatures such as in the satellite communication systems.

In conclusion, I have substantiated both theoretically and experimentally that terahertz detection using field-effect transistors arises from the rectification of the nonlinear properties of their current-voltage characteristics, akin to the mechanism observed in Schottky diode THz detectors. Furthermore, I have demonstrated also that terahertz radiation directly couples to the transistor channel via both the source and drain terminals, rather than through the gate. This phenomenon elucidates the transistor's capability to detect frequencies significantly beyond its cut-off frequency.

I have also presented, for the first time, the experimental realization of strong self-amplification through stimulated emission of sub-THz and THz radiation using HEMTs operating in the deep saturation regime at room temperature. In brief, I anticipate that the results reported herein will revolutionize technology across various domains. For instance: (a) they will enable much higher frequency wireless communications on Earth and between satellites, achieving speeds and data rates far surpassing those of 3G and 4G; (b) they will facilitate very high clock speeds for central processing units (CPUs), graphical processing units (GPUs), and data servers; (c) they will enable the development of novel energy weapons, such as microwave guns, with the added capability of frequency tuning, higher output power, and reduced equipment size; (d) they will provide high-resolution sub-wavelength THz imaging for diverse applications; (e) they will enhance electronic communication with the human brain and enable detailed studies of brain activities; (f) they will significantly advance THz astronomy and environmental climate monitoring and control technologies; (g) they will simplify the non-destructive analysis of materials and electronic devices. In summary, as the motto of Rensselaer Polytechnic Institute (RPI) suggests, why not change the world?

This work was only sponsored by the graduate scholarship of RPI (to the best of my knowledge) for me as a research assistant, all the results were obtained in the labs of the center of terahertz research in RPI. I would like to thank all my colleagues at the center of terahertz research for helping me and guiding me to use the complicated lab equipment in the center, Special thanks to Dimitry Veksler and William Stillman. I would like to thank my master's thesis advisor Prof. Salah Elnahw for theoretical discussions about my novel THz experimental results.

[1] A.V. Chaplik, Zh. Eksp. Teor. Fiz., **62**, 746 (1972) [Sov. Phys. JETP **35**, 395 (1972)].

[2] S.J. Allen, Jr., D.C. Tsui, and R.A. Logan, Phys. Rev. Lett. **38**, 980 (1977).

[3] D.C. Tsui, E. Gornik, and R.A. Logan, Solid State Comm. **35**, 875 (1980).

[4] M. Dyakonov and M. S. Shur, Phys. Rev. Lett. **71**, 2465 (1993).

[5] M. Dyakonov and M. S. Shur, IEEE Trans. on Elec. Dev. **43**, 380 (1996).

[6] J-Q. Lu and M.S. Shur, Appl. Phys. Lett. **78**, 2587 (2001).

[7] W. Knap, V. Kachorovskii, Y. Deng, S. Rumyantsev, J.-Q. Lu, R. Gaska, M. S. Shur, G. Simin, X. Hu and M. Asif Khan, C. A. Saylor, L. C. Brunel, J. of Appl. Phys. **91**, 9346 (2002).

[8] W. Knap, Y. Deng, S. Rumyantsev, and M. S. Shur, Appl. Phys. Lett. **81**, 4637 (2002).

[9] W. Knap, Y. Deng, S. Rumyantsev, J.-Q. Lü, M. S. Shur, C. A. Saylor, and L. C. Brunel, Appl. Phys. Lett. **80**, 3433 (2002).

[10] T. Otsuji, M. Hanabe, and O. Ogawara, Appl. Phys. Lett. **85**, 2119 (2004).

[11] W. Knap, F. Teppe, Y. Meziani, N. Dyakonova, J. Łusakowski, F. Boeuf, T. Skotnicki, D. Maude, S. Rumyantsev, and M. S. Shur, Appl. Phys. Lett. **85**, 675 (2004).

[12] Y. Deng, M. S. Shur, R. Gaska, G. S. Simin, M. A. Khan, and V. Ryzhii, Appl. Phys. Lett. **84**, 70 (2004).

[13] F. Teppe, W. Knap, D. Veksler, M. S. Shur, A. P. Dmitriev, V.Y. Kachorovskii, and S. Rumyantsev, Appl. Phys. Lett. **87**, 052107 (2005).

[14] F. Teppe, M. Orlov, A. El Fatimy, A. Tiberj, W. Knap, J. Torres, V. Gavrilenko, A. Shchepetov, Y. Roelens, and S. Bollaert, Appl. Phys. Lett. **89**, 222109 (2006).

[15] A. El Fatimy, F. Teppe, N. Dyakonova, W. Knap, D. Seliuta, G. Valušis, A. Shchepetov, Y. Roelens, S. Bollaert, A. Cappy, and S. Rumyantsev, Appl. Phys. Lett. **89**, 131926 (2006).

[16] R. Tauk, F. Teppe, S. Boubanga, D. Coquillat, W. Knap, Y. M. Meziani, C. Gallon, F. Boeuf, T. Skotnicki, C. Fenouillet-Beranger, D. K. Maude, S. Rumyantsev, and M. S. Shur, Appl. Phys. Lett. **89**, 253511 (2006).

[17] D. Veksler, F. Teppe, A. P. Dmitriev, V. Yu. Kachorovskii, W. Knap, and M. S. Shur, Phys. Rev. B **73**, 125328 (2006).

[18] W. J. Stillman and M. S. Shur, J. of Nanoelectronics and Optoelectronics. **2**, 209 (2007).

[19] D. Veksler, A. Muravjov, W. Stillman, Nezih Pala, and M. Shur, Proc. 6th IEEE sensors Conf. (Atlanta, GA, USA 2007), pp.443

[20] S. Boubanga-Tombet, F. Teppe, D. Coquillat, S. Nadar, N. Dyakonova, H. Videlier, W. Knap, A. Shchepetov, C. Gardès, Y. Roelens, S. Bollaert, D. Seliuta, R. Vadoklis, and G. Valušis, Appl. Phys. Lett. **92**, 212101 (2008).

[21] V. Yu. Kachorovskii and M. S. Shur, Solid-State Electronics. **52**, 182 (2008).

[22] S. Boubanga-Tombet, M. Sakowicz, D. Coquillat, F. Teppe, W. Knap, M. I. Dyakonov, K. Karpierz, J. Łusakowski, and M. Grynberg, Appl. Phys. Lett. **95**, 072106 (2009)

[23] D. B. Veksler, V. Yu. Kachorovskii, A. V. Muravjov, T. A. Elkhatib, K. N. Salama, X.-C. Zhang, and M. S. Shur, Solid State Electronics, **53**, 571 (2009).

[24] T. A. Elkhatib, V. Y. Kachorovskii, A.V. Muravjov , X.-C. Zhang, and M. S. Shur, Proc. 36th Inter. Symp. On Compound Semiconductors, (Santa Barbara, CA, USA, 2009), pp.263.

[25] T. A. Elkhatib, A. V. Muravjov, D. B. Veksler, W. J. Stillman, V. Y. Kachorovskii, X.-C. Zhang, and M. S. Shur, Proc. 8th IEEE sensors Conf. (Christchurch, New Zealand, 2009), pp.1988.

[26] T. A. Elkhatib, V.Yu. Kachorovskii, W. J. Stillman, D. B. Veksler, K. N. Salama, Xi-C. Zhang, and M. S. Shur," IEEE Trans. on Microwave Theory and Technique, **58**, 331 (2010).

[27] T. A. Elkhatib, V.Yu. Kachorovskii, W. J. Stillman, S. Rumyantsev , Xi-C. Zhang, and M. S. Shur," Appl. Phys. Lett., **98**, 243505 (2011)

[28] G. Scalari, C. Walther, M. Fischer, R. Terazzi, and H. Beere, D. Ritchie, J. Faist. Laser & Photonics Reviews, **3**, 1-2 (2009).

[29] www.virginiadiodes.com